\documentclass[final]{ws-procs9x6}

\newcommand{\hepph}[1]{{\tt hep-ph/#1}}
\newcommand{\hepex}[1]{{\tt hep-ex/#1}}
\newcommand{\astroph}[1]{{\tt astro-ph/#1}}

\def\beq{\begin{equation}}
\def\eeq{\end{equation}}
\def\bea{\begin{eqnarray}}
\def\eea{\end{eqnarray}}
\def\openep{\leavevmode\hbox{\normalsize$\epsilon$
\kern-4.pt$\epsilon$}}

\begin{document}

\title{Yukawa Quasi-Unification and Inflation}

\author{G. Lazarides and C. Pallis}

\address{Physics Division, School of Technology,\\
Aristotle University of Thessaloniki,\\
54124 Thessaloniki, GREECE \\
E-mail: lazaride@eng.auth.gr, kpallis@auth.gr }

\maketitle

\pagestyle{headings}

\thispagestyle{plain}\markboth{G. Lazarides and
C. Pallis}{Yukawa Quasi-Unification and Inflation}
\setcounter{page}{1}

\abstracts{We review the construction of a concrete
supersymmetric grand unified model, which naturally
leads to a moderate violation of `asymptotic' Yukawa
unification and, thus, can allow an acceptable
$b$-quark mass within the constrained minimal
supersymmetric standard model with $\mu>0$. The
model possesses a wide and natural range of
parameters which is consistent with the data on the
cold dark matter abundance in the universe,
$b\rightarrow s\gamma$, the muon anomalous magnetic
moment and the Higgs boson masses. Also, it
automatically leads to a new version of shifted
hybrid inflation, which avoids overproduction of
monopoles at the end of inflation by using only
renormalizable terms.}

\section{Introduction} \label{sec:intro}

The most restrictive version of the minimal
supersymmetric standard model (MSSM) with gauge
coupling unification, radiative
electroweak breaking and universal boundary
conditions from gravity-mediated soft
supersymmetry (SUSY) breaking, known as
constrained MSSM (CMSSM)\cite{Cmssm}, can be
made even more predictive, if we impose Yukawa
unification (YU), i.e. assume that
the three third generation Yukawa coupling
constants unify at the
SUSY grand unified theory (GUT) scale,
$M_{\rm GUT}$. The requirement of YU can be
achieved by embedding the MSSM in a SUSY
GUT with a gauge group containing
${\rm SU(4)}_c$ and ${\rm SU(2)}_R$. Indeed,
assuming that the electroweak Higgs superfields
$h^{\rm ew}_1$, $h^{\rm ew}_2$ and the third
family right handed quark superfields $t^c$,
$b^c$ form ${\rm SU(2)}_R$ doublets, we
obtain\cite{pana} the `asymptotic' Yukawa
coupling relation $h_t=h_b$ and, hence, large
$\tan\beta\sim m_{t}/m_{b}$. Moreover,
if the third generation quark and lepton
${\rm SU(2)}_L$ doublets [singlets] $q_3$ and
$l_3$ [$b^c$ and $\tau^c$] form a ${\rm SU(4)}_c$
{\bf 4}-plet [${\bf\bar 4}$-plet] and the Higgs
doublet $h^{\rm ew}_1$ which couples to them is
a ${\rm SU(4)}_c$ singlet, we get $h_b=h_{\tau}$
and the `asymptotic' relation $m_{b}=m_{\tau}$
follows. The simplest GUT gauge group which
contains both ${\rm SU(4)}_c$ and ${\rm SU(2)}_R$
is the Pati-Salam (PS) group $G_{\rm PS}=
{\rm SU(4)}_c\times {\rm SU(2)}_L
\times {\rm SU(2)}_R$ and we will use it in our
analysis.

\par
However, applying YU in the context of the CMSSM and
given the experimental values of the top-quark and
tau-lepton masses (which naturally restrict
$\tan\beta\simeq50$), the resulting value of
the $b$-quark mass turns out to be unacceptable. This
is due to the fact that, in the large $\tan\beta$
regime, the tree-level $b$-quark mass receives
sizeable SUSY corrections\cite{copw,pierce,susy,king}
(about 20$\%$), which have the sign of $\mu$ (with
the standard sign convention\cite{sugra}) and drive,
for $\mu>[<]~0$, the corrected $b$-quark mass at
$M_Z$, $m_b(M_Z)$, well above [somewhat below] its
$95\%$ confidence level (c.l.) experimental range:
\begin{equation} 2.684~{\rm GeV}\lesssim m_b(M_Z)
\lesssim3.092~{\rm
GeV}~~\mbox{with}~~\alpha_s(M_Z)=0.1185.
\label{mbrg}
\end{equation}
This is derived by appropriately\cite{qcdm} evolving
the corresponding range of $m_b(m_b)$ in the
$\overline{MS}$ scheme (i.e. $3.95-4.55~{\rm GeV}$)
up to $M_{Z}$ in accordance with Ref.~\refcite{baermb}.
We see that, for both signs of $\mu$, YU leads to an
unacceptable $b$-quark mass with the $\mu<0$ case
being less disfavored.

\par
A way out of this $m_b$ problem can be found\cite{qcdm}
without abandoning the CMSSM (in contrast to the usual
strategy\cite{king,raby,baery,nath}) or YU altogether.
We can rather modestly correct YU by including an extra
${\rm SU(4)}_c$ non-singlet Higgs superfield with
Yukawa couplings to the quarks and leptons. The Higgs
${\rm SU(2)}_L$ doublets contained in this superfield
can naturally develop\cite{wetterich} subdominant
vacuum expectation values (VEVs) and mix with the
main electroweak doublets, which are assumed to be
${\rm SU(4)}_c$ singlets and form a ${\rm SU(2)}_R$
doublet. This mixing can, in general, violate
${\rm SU(2)}_R$. Consequently, the resulting
electroweak Higgs doublets $h_1^{\rm ew}$,
$h_2^{\rm ew}$ do not form a ${\rm SU(2)}_R$ doublet
and also break ${\rm SU(4)}_c$. The required
deviation from YU is expected to be more pronounced
for $\mu>0$. Despite this, we will study here this
case, since the $\mu<0$ case has been
excluded\cite{cd2} by combining the Wilkinson
microwave anisotropy probe (WMAP)
restrictions\cite{wmap} on the cold dark matter (CDM)
in the universe with the recent experimental
results\cite{cleo} on the inclusive branching ratio
${\rm BR}(b\rightarrow s\gamma)$. The same SUSY GUT
model which, for $\mu>0$ and universal boundary
conditions, remedies the $m_b$ problem leads to a
new version\cite{jean2} of shifted hybrid
inflation\cite{jean}, which avoids monopole
overproduction at the end of inflation and, in
contrast to the older version\cite{jean},
is based only on renormalizable interactions.

\par
In Sec.~\ref{model}, we review the construction of a
SUSY GUT model which modestly violates YU, yielding
an appropriate Yukawa quasi-unification condition
(YQUC), which is derived in Sec.~\ref{yquc}. We then
describe the resulting CMSSM in Sec.~\ref{qcmssm}
and introduce the various cosmological and
phenomenological requirements which restrict its
parameter space in Sec.~\ref{pheno}. In
Sec.~\ref{parameters}, we delineate the allowed
range of parameters for $\mu>0$ and, in
Sec.~\ref{inflation}, we outline the corresponding
inflationary scenario. Finally, in
Sec.~\ref{conclusions}, we summarize our conclusions.

\section{The SUSY GUT Model} \label{model}

We will take the SUSY GUT model of shifted hybrid
inflation\cite{jean}
(see also Ref.~\refcite{talks}) as our starting point.
It is based on $G_{\rm PS}$, which is the simplest
gauge group that can lead to YU. The representations
under $G_{\rm PS}$ and the global charges of the
various matter and Higgs superfields contained in
this model are presented in Table 1, which also
contains the extra Higgs superfields required for
accommodating an adequate violation of YU (see below).
The matter superfields are $F_i$ and $F^c_i$
($i=1,2,3$), while the electroweak Higgs doublets
belong to the superfield $h$. So, all the requirements
for exact YU are fulfilled. The breaking of
$G_{\rm PS}$ down to the standard model (SM) gauge
group $G_{\rm SM}$ is achieved by the superheavy VEVs
($\sim M_{\rm GUT}$) of the right handed neutrino type
components of a conjugate pair of Higgs superfields
$H^c$, $\bar{H}^c$. The model also contains a gauge
singlet $S$ which triggers the breaking of $G_{\rm PS}$,
a ${\rm SU(4)}_c$ {\bf 6}-plet $G$ which
gives\cite{leontaris} masses to the right handed down
quark type components of $H^c$, $\bar{H}^c$, and a pair
of gauge singlets $N$, $\bar{N}$ for solving\cite{rsym}
the $\mu$ problem of the MSSM via a Peccei-Quinn (PQ)
symmetry. In addition to $G_{\rm PS}$, the model
possesses two global ${\rm U(1)}$ symmetries,
namely a R and a PQ symmetry, as well as a discrete
$Z_2^{\rm mp}$ symmetry (`matter parity').

\par
A moderate violation of exact YU can be naturally
accommodated in this model by adding a new Higgs
superfield $h^{\prime}$ with Yukawa couplings
$FF^ch^{\prime}$. Actually, ({\bf 15,2,2}) is
the only representation, besides ({\bf 1,2,2}),
which possesses such couplings to the fermions.
In order to give superheavy masses to the color
non-singlet components of $h^{\prime}$, we need
to include one more Higgs superfield
$\bar{h}^{\prime}$ with the superpotential
coupling $\bar{h}^{\prime}h^{\prime}$, whose
coefficient is of the order of $M_{\rm GUT}$.

\begin{table}[!t]

\tbl{Superfield Content of the Model}
{\footnotesize
\begin{tabular}{@{}ccccc@{}}
\hline {} &{} &{} &{} &{}\\[-1.5ex]
{Superfields}&{Representations}&
\multicolumn{3}{c}{Global}
\\[1ex] \multicolumn{1}{c}{}&{under $G_{\rm PS}$}
&\multicolumn{3}{c}{Symmetries}
\\[1ex] {}&{}&{$R$} &{$ PQ$} &{$Z^{\rm mp}_2$}
\\[1ex] \hline
{} &{} &{} &{} &{}\\[-1.5ex]
\multicolumn{5}{c}{Matter Fields}
\\[1ex] \hline
{} &{} &{} &{} &{}\\[-1.5ex]
{$F_i$} &{$({\bf 4, 2, 1})$}& $1/2$ & $-1$ &$1$
 \\[1ex]
{$F^c_i$} & {$({\bf \bar 4, 1, 2})$} &{ $1/2$ }
&{$0$}&{$-1$}
\\[1ex] \hline {} &{} &{} &{} &{}\\[-1.5ex]
\multicolumn{5}{c}{Higgs Fields}
\\[1ex] \hline {} &{} &{} &{} &{}\\[-1.5ex]
{$h$} & {$({\bf 1, 2, 2})$}&$0$ &$1$ &$0$\\[1ex]
\hline {} &{} &{} &{} &{}\\[-1.5ex]
{$H^c$} &{$({\bf \bar 4, 1, 2})$}&{$0$}&{$0$}
& {$0$}
\\[1ex]
{$\bar H^c$}&$({\bf 4, 1, 2})$&{$0$}&{$0$}&{$0$}
\\[1ex]
{$S$} & {$({\bf 1, 1, 1})$}&$1$ &$0$ &$0$ \\[1ex]
{$G$} & {$({\bf 6, 1, 1})$}&$1$ &$0$ &$0$\\[1ex]
\hline {} &{} &{} &{}
&{}\\[-1.5ex]
{$N$} &{$({\bf 1, 1, 1})$}& {$1/2$}&{$-1$} & {$0$}
\\[1ex]
{$\bar N$}&$({\bf 1, 1, 1})$& {$0$}&{$1$}&{$0$}
\\ [1ex]\hline {} &{} &{} &{} &{}\\[-1.5ex]
\multicolumn{5}{c}{Extra Higgs Fields}
\\[1ex] \hline {} &{} &{} &{} &{}\\[-1.5ex]
$h^{\prime}$&{$({\bf 15, 2, 2})$}& $0$ & $1$ &$0$
\\[1ex]
$\bar h^{\prime}$&{$({\bf 15, 2, 2})$}& $1$ & $-1$
&$0$
\\[1ex]
$\phi$&$({\bf 15, 1, 3})$ & $0$ & $0$ &$0$
\\[1ex]
$\bar\phi$&{$({\bf 15, 1, 3})$}  & $1$ & $0$ &$0$
\\[1ex]\hline
\end{tabular}\label{table} }
\vspace*{-13pt}
\end{table}

\par
After the breaking of $G_{\rm PS}$ to $G_{\rm SM}$,
the two color singlet ${\rm SU(2)}_L$ doublets
$h_1^{\prime}$, $h_2^{\prime}$ contained in
$h^{\prime}$ can mix with the corresponding doublets
$h_1$, $h_2$ in $h$. This is mainly due to the terms
$\bar{h}^{\prime}h^{\prime}$ and
$H^c\bar{H}^c\bar{h}^{\prime}h$. Actually, since
\begin{eqnarray} && \nonumber
H^c\bar H^c =({\bf\bar 4,1,2})({\bf 4,1,2})=
({\bf 15, 1, 1+3})+\cdots,\\ &&\nonumber
\bar h^\prime h=({\bf 15,2,2})
({\bf 1,2,2})=({\bf 15, 1, 1+3})+\cdots,
\end{eqnarray}
there are two independent couplings of the type
$H^c\bar{H}^c\bar{h}^{\prime}h$ (both suppressed by
the string scale
$M_{\rm S}\approx 5\times 10^{17}~{\rm GeV}$, being
non-renormalizable). One of them is between the
${\rm SU(2)}_R$ singlets in $H^c\bar{H}^c$ and
$\bar{h}^{\prime}h$, and the other between the
${\rm SU(2)}_R$ triplets in these combinations. So,
we obtain two bilinear terms $\bar{h}_1^{\prime}h_1$
and $\bar{h}_2^{\prime}h_2$ with different
coefficients, which are suppressed by
$M_{\rm GUT}/M_{\rm S}$. These terms together with
the terms $\bar{h}_1^{\prime}h_1^{\prime}$ and
$\bar{h}_2^{\prime}h_2^{\prime}$ from
$\bar{h}^{\prime}h^{\prime}$, which have equal
coefficients, generate different mixings between
$h_1$, $h_1^{\prime}$ and $h_2$, $h_2^{\prime}$.
Consequently, the resulting electroweak doublets
$h_1^{\rm ew}$, $h_2^{\rm ew}$ contain
${\rm SU(4)}_c$ violating components suppressed by
$M_{\rm GUT}/M_{\rm S}$ and fail to form a
${\rm SU(2)}_R$ doublet by an equally suppressed
amount. So, YU is moderately violated.
Unfortunately, as it turns out, this violation is
not adequate for correcting the $b$-quark mass
within the CMSSM for $\mu>0$.

\par
In order to allow for a more sizable violation of YU,
we further extend the model by including $\phi$ with
the coupling $\phi\bar{h}^{\prime}h$. To give
superheavy masses to the color non-singlets in $\phi$,
we introduce one more superfield $\bar\phi$ with the
coupling $\bar\phi\phi$, whose coefficient
is of order $M_{\rm GUT}$.

\par
The terms $\bar\phi\phi$ and $\bar\phi H^c\bar{H}^c$
imply that, after the breaking of $G_{\rm PS}$ to
$G_{\rm SM}$, $\phi$ acquires a superheavy VEV of
order $M_{\rm GUT}$. The coupling
$\phi\bar{h}^{\prime}h$ then generates
${\rm SU(2)}_R$ violating unsuppressed bilinear
terms between the doublets in $\bar{h}^{\prime}$
and $h$. These terms can certainly overshadow
the corresponding ones from the non-renormalizable
term $H^c\bar{H}^c\bar{h}^{\prime}h$. The resulting
${\rm SU(2)}_R$ violating mixing of the doublets in
$h$ and $h^{\prime}$ is then unsuppressed and we can
obtain stronger violation of YU.

\section{The Yukawa Quasi-Unification Condition}
\label{yquc}

To further analyze the mixing of the doublets in $h$
and $h^{\prime}$, observe that the part of the
superpotential corresponding to the symbolic couplings
$\bar{h}^{\prime}h^{\prime}$, $\phi\bar{h}^{\prime}h$
is properly written as
\begin{equation}
m{\rm tr}\left(\bar h^{\prime}\epsilon \tilde
h^{\prime}\epsilon\right)+p{\rm tr}\left(
\bar h^{\prime}
\epsilon\phi \tilde h\epsilon \right),
\label{expmix}
\end{equation}
where $\openep$ is the antisymmetric $2\times2$
matrix with $\epsilon_{12}=+1$, ${\rm tr}$ denotes
trace taken with respect to the ${\rm SU(4)}_{\rm c}$
and ${\rm SU(2)}_L$ indices and tilde denotes the
transpose of a matrix.

\par
After the breaking of $G_{\rm PS}$ to $G_{\rm SM}$,
$\phi$ acquires a VEV
$\langle\phi \rangle\sim M_{\rm GUT}$. Substituting
it by this VEV in the above couplings, we obtain
\begin{eqnarray}
&& {\rm tr}(\bar{h}^{\prime}\epsilon\tilde{h}^{\prime}
\epsilon)=\tilde{\bar{h}}^{\prime}_1\epsilon
h^{\prime}_2+\tilde{h}^{\prime}_1\epsilon
\bar{h}^{\prime}_2+\cdots,  \label{mass}
\\
&& {\rm tr}(\bar{h}^{\prime}\epsilon\phi\tilde{h}
\epsilon)=\frac{\langle\phi\rangle}{\sqrt{2}}{\rm tr}
(\bar{h}^{\prime}\epsilon\sigma_3\tilde{h}\epsilon)=
\frac{\langle\phi\rangle}{\sqrt{2}}
(\tilde{\bar{h}}^{\prime}_1\epsilon h_2-\tilde{h}_1
\epsilon\bar{h}^{\prime}_2), \label{triplet}
\end{eqnarray}
where the ellipsis in Eq.~(\ref{mass}) contains the
colored components of $\bar{h}^{\prime}$,
$h^{\prime}$ and $\sigma_3={\rm diag}(1,-1)$.
Inserting Eqs.~(\ref{mass}) and (\ref{triplet}) into
Eq. (\ref{expmix}), we obtain
\begin{equation}
m\tilde{\bar h}^\prime_1\epsilon(h^{\prime}_2-
\alpha_1h_2)+m(\tilde{h}^{\prime}_1+\alpha_1
\tilde{h}_1)\epsilon\bar{h}^{\prime}_2~~\mbox{with}~~
\alpha_1=-p\langle\phi\rangle/\sqrt{2}m.\quad
\label{superheavy}
\end{equation}
So, we get two pairs of superheavy doublets with mass
$m$. They are predominantly given by
\begin{equation}
\bar{h}^{\prime}_1~,~\frac{h^{\prime}_2-\alpha_1h_2}
{\sqrt{1+|\alpha_1|^2}}~~{\rm and}~~
\frac{h^{\prime}_1+\alpha_1h_1}
{\sqrt{1+|\alpha_1|^2}}~,~\bar{h}^{\prime}_2.
\label{superdoublets}
\end{equation}
The orthogonal combinations of $h_1$, $h^{\prime}_1$
and $h_2$, $h^{\prime}_2$ constitute the electroweak
doublets
\begin{equation}
h_1^{\rm ew}=\frac{h_1-\alpha_1^*h^{\prime}_1}
{\sqrt{1+|\alpha_1|^2}}~~{\rm and}~~ h_2^{\rm
ew}=\frac{h_2+\alpha_1^*h^{\prime}_2}
{\sqrt{1+|\alpha_1|^2}}\cdot
\label{ew}
\end{equation}
The superheavy doublets in Eq.~(\ref{superdoublets})
must have vanishing VEVs, which readily implies that
$\langle h_1^{\prime} \rangle=-\alpha_1\langle
h_1\rangle$, $\langle h_2^{\prime}\rangle=\alpha_1
\langle h_2\rangle$. Equation (\ref{ew}) then gives
$\langle h_1^{\rm ew}\rangle=(1+|\alpha_1|^2)^{1/2}
\langle h_1 \rangle$, $\langle h_2^{\rm ew}\rangle=
(1+|\alpha_1|^2) ^{1/2}\langle h_2\rangle$.
From the third generation Yukawa couplings
$y_{33}F_3hF_3^c$,
$2y_{33}^{\prime}F_3h^{\prime}F_3^c$, we obtain
\begin{eqnarray}
&& m_t=|y_{33}\langle h_2\rangle+y_{33}^{\prime}
\langle
h_2^{\prime}\rangle|=\left|\frac{1+\rho\alpha_1/
\sqrt{3}}{\sqrt{1+|\alpha_2|^2}}y_{33} \langle
h_2^{\rm ew}\rangle\right|,\label{top} \\ &&
m_b=\left|\frac{1-\rho\alpha_1/
\sqrt{3}}{\sqrt{1+|\alpha_1|^2}}y_{33} \langle
h_1^{\rm ew}\rangle\right|,~ m_\tau=
\left|\frac{1+\sqrt{3}\rho\alpha_1}
{\sqrt{1+|\alpha_1|^2}}y_{33} \langle
h_1^{\rm ew}\rangle\right|\cdot
\label{bottomtau}
\end{eqnarray}
where $\rho=y_{33}^{\prime}/y_{33}$. From
Eqs.~(\ref{top}) and (\ref{bottomtau}), we see that
YU is now replaced by the YQUC,
\begin{equation}
h_t:h_b:h_\tau=(1+c):(1-c):(1+3c),~~\mbox{with}~~
0<c=\rho\alpha_1/\sqrt{3}<1.
\label{minimal}
\end{equation}
For simplicity, we restricted ourselves to real
values of $c$ only which lie between zero and unity.

\section{The resulting CMSSM}\label{qcmssm}

Below $M_{\rm GUT}$, the particle content of our
model reduces to this of MSSM (modulo SM singlets).
We assume universal soft SUSY breaking scalar masses
$m_0$, gaugino masses $M_{1/2}$ and
trilinear scalar couplings $A_0$ at $M_{\rm GUT}$.
Therefore, the resulting MSSM is the so-called
CMSSM\cite{Cmssm} with $\mu>0$ and supplemented by
Eq.~(\ref{minimal}). With these initial conditions,
we run the MSSM renormalization group equations
(RGEs)\cite{cdm} between $M_{\rm GUT}$ and a common
SUSY threshold $M_{\rm SUSY}\simeq(m_{\tilde t_1}
m_{\tilde t_2})^{1/2}$ ($\tilde t_{1,2}$
are the stop mass eigenstates) determined in
consistency with the SUSY spectrum. At
$M_{\rm SUSY}$, we impose radiative electroweak
symmetry breaking, evaluate the SUSY spectrum and
incorporate the SUSY
corrections\cite{pierce,susy,king} to the $b$ and
$\tau$ masses. Note that the corrections to the
$\tau$ mass (almost 4$\%$) lead\cite{cd2} to a small
reduction of $\tan\beta$. From $M_{\rm SUSY}$ to
$M_Z$, the running of gauge and Yukawa coupling
constants is continued using the SM RGEs.

\par
For presentation purposes, $M_{1/2}$ and $m_0$ can be
replaced\cite{cdm} by the lightest SUSY particle (LSP)
mass, $m_{\rm LSP}$, and the relative mass splitting
between this particle and the lightest stau
$\tilde\tau_2$, $\Delta_{\tilde\tau_2}=(m_{\tilde\tau_2}
-m_{\rm LSP })/m_{\rm LSP}$. For simplicity, we restrict
this presentation to the $A_0=0$ case (for $A_0\neq0$
see Refs.~\refcite{qcdm} and \refcite{mario}). So, our
input parameters are $m_{\rm LSP}$ and
$\Delta_{\tilde\tau_2}$.

\par
For any given $m_b(M_Z)$ in the range in
Eq.~(\ref{mbrg}) and with fixed $m_t(m_t)=166~{\rm GeV}$
and $m_\tau(M_Z)=1.746~{\rm GeV}$, we can determine the
parameters $c$ and $\tan\beta$ at $M_{\rm SUSY}$ so that
the YQUC in Eq.~(\ref{minimal}) is satisfied.

\section{Cosmological and Phenomenological Constraints}
\label{pheno}

Restrictions on the parameters of our model can be
derived by
imposing a number of cosmological and phenomenological
requirements (for similar recent analyses, see
Refs.~\refcite{baery,nath} and \refcite{spanos}).
These constraints result from:

$\bullet$ {\em Cold dark matter considerations}. In
the context of CMSSM, the LSP can be the lightest
neutralino. It naturally arises\cite{goldberg} as a
CDM candidate. We require its relic abundance,
$\Omega_{\rm LSP}h^2$, not to exceed the $95\%$ c.l.
upper bound on the CDM abundance derived\cite{wmap}
by WMAP:
\begin{equation}
\Omega_{\rm CDM}h^2\lesssim0.13.\label{cdmb}
\end{equation}
We calculate $\Omega_{\rm LSP}h^2$ using {\tt
micrOMEGAs}\cite{micro}, which is certainly one of
the most complete publicly available codes. It
includes all possible coannihilation
processes\cite{ellis} and one-loop QCD
corrections\cite{width} to the Higgs decay widths
and couplings to fermions.

$\bullet$ {\em Branching ratio of
$b\rightarrow s\gamma$}. Taking
into account the recent experimental
results\cite{cleo} on this ratio,
${\rm BR}(b\rightarrow s\gamma)$, and
combining\cite{qcdm} appropriately the experimental
and theoretical errors involved, we obtain the
following $95\%$ c.l. range:
\begin{equation} 1.9\times 10^{-4}\lesssim
{\rm BR}(b\rightarrow
s\gamma)\lesssim 4.6 \times 10^{-4}.
\label{bsgb} \end{equation}
We compute ${\rm BR}(b\rightarrow s\gamma)$ by using
an updated version of the relevant calculation
contained in the {\tt micrOMEGAs} package\cite{micro}.
In this code, the SM contribution
is calculated following Ref.~\refcite{kagan}. The
charged Higgs ($H^\pm$) contribution is evaluated by
including the next-to-leading order (NLO) QCD
corrections\cite{nlo} and $\tan\beta$ enhanced
contributions\cite{nlo}. The dominant SUSY
contribution includes resummed NLO SUSY QCD
corrections\cite{nlo}, which hold for large
$\tan\beta$.

$\bullet$ {\em Muon anomalous magnetic moment}. The
deviation, $\delta a_\mu$, of the measured value of
$a_\mu$ from its predicted value in the SM,
$a^{\rm SM}_\mu$, can be attributed to SUSY
contributions, calculated by using the
{\tt micrOMEGAs} routine\cite{gmuon}. The
calculation of $a^{\rm SM}_\mu$ is not
yet stabilized mainly because of the instability of
the hadronic vacuum polarization contribution.
According to the most up-to-date
evaluation\cite{davier}, there is still a
considerable discrepancy between the findings based
on the $e^+e^-$ annihilation data and the ones
based on the $\tau$-decay data. Taking into account
these results and the experimental
measurement\cite{muon} of $a_\mu$, we
get the following $95\%$ c.l. ranges:
\begin{eqnarray} -0.53\times10^{-10}\lesssim&\delta
a_\mu&\lesssim 44.7\times
10^{-10},~\quad\mbox{$e^+e^-$-based};\label{g2e}
\\\vspace*{19pt}
-13.6\times10^{-10}\lesssim&\delta a_\mu&\lesssim28.4
\times 10^{-10},\quad\mbox{$\tau$-based}.
\label{g2t}
\end{eqnarray}
Following the common practice\cite{spanos}, we adopt the
restrictions to parameters induced by Eq. (\ref{g2e}),
since Eq.~(\ref{g2t}) is considered as quite oracular,
due to poor $\tau$-decay data.

$\bullet$ {\em Collider bounds}. Here, the only relevant
collider bound is the $95\%$ c.l. LEP bound\cite{higgs}
on the mass of the lightest CP-even neutral Higgs boson
$h$:
\begin{equation}
m_h\gtrsim114.4~{\rm GeV}.
\label{mhb}
\end{equation}
The SUSY corrections to $m_h$ are calculated at two
loops by using the {\tt FeynHiggsFast}
program\cite{fh} included
in the {\tt micrOMEGAs} code\cite{micro}.

\begin{figure}[t]
\centerline{\begin{turn}{-90}\epsfxsize=2.7in
\epsfbox{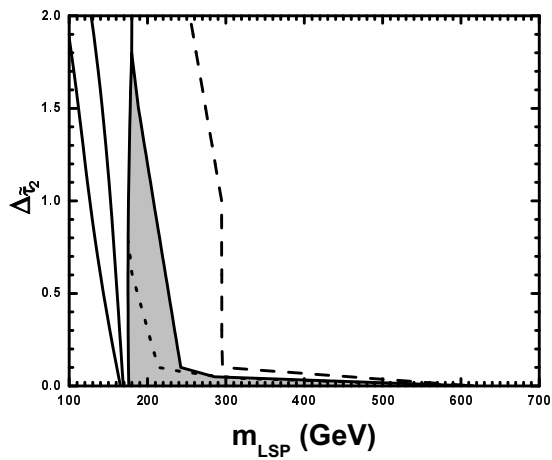}\end{turn}}
\caption{The various restrictions on the
$m_{\rm LSP}-\Delta_{\tilde\tau_2}$
plane for $\mu>0$, $A_0=0$ and $\alpha_s(M_Z)=0.1185$.
From left to right, the solid lines depict the lower
bounds on $m_{\rm LSP}$ from $\delta a_\mu<44.7
\times 10^{-10}$, ${\rm BR}(b\rightarrow
s\gamma)>1.9\times 10^{-4}$ and
$m_h>114.4~{\rm GeV}$ and the upper bound on
$m_{\rm LSP}$ from $\Omega_{\rm LSP}h^2<0.13$ for
$m_b(M_Z)=2.888~{\rm GeV}$. The dashed [dotted] line
depicts the bound on $m_{\rm LSP}$ from
$\Omega_{\rm LSP}h^2<0.13$ for
$m_b(M_Z)=2.684~[3.092]~{\rm GeV}$. The allowed area
for $m_b(M_Z)=2.888~{\rm GeV}$ is shaded.
\label{figa}}
\end{figure}

\section{The Allowed Parameter Space}
\label{parameters}

We will now try to delineate the parameter space
of our model with $\mu>0$ which is consistent with
the constraints in Sec.~\ref{pheno}. The restrictions
on the $m_{\rm LSP}-\Delta_{\tilde\tau_2}$ plane, for
$A_0=0$ and the central values of $\alpha_s(M_Z)=0.1185$
and $m_b(M_Z)=2.888~{\rm GeV}$, are indicated in
Fig.~\ref{figa} by solid lines, while the upper bound on
$m_{\rm LSP}$ from Eq.~(\ref{cdmb}), for
$m_b(M_Z)=2.684~[3.092]~{\rm GeV}$, is depicted by a
dashed [dotted] line. We observe the following:

\begin{figure}[ht]
\centerline{\begin{turn}{-90}\epsfxsize=2.7in
\epsfbox{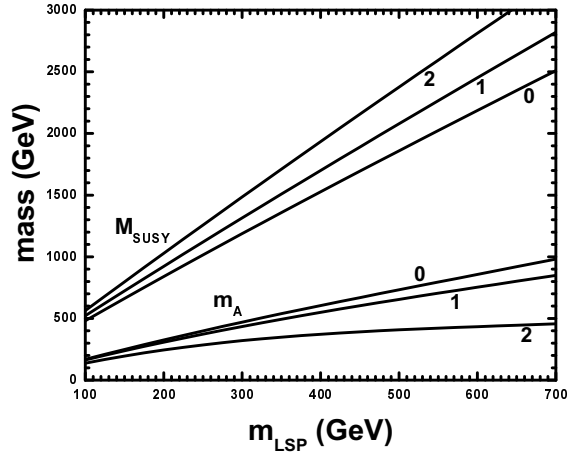}\end{turn}}
\caption{The mass parameters $m_A$ and $M_{\rm SUSY}$
as functions of $m_{\rm LSP}$ for various values of
$\Delta_{\tilde\tau_2}$, which are indicated on the
curves. We take $\mu>0$, $A_0=0$,
$m_b(M_Z)=2.888~{\rm GeV}$ and $\alpha_s(M_Z)=0.1185$.
\label{figb}}
\end{figure}

\begin{figure}[hb]
\centerline{\epsfxsize=2.7in\begin{turn}{-90}
\epsfbox{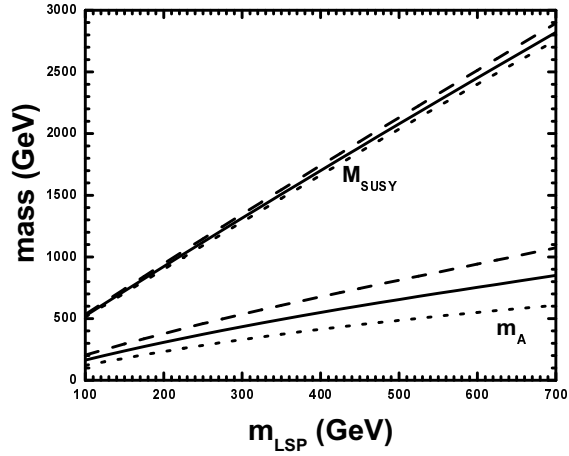}\end{turn}}
\caption{The mass parameters $m_A$ and $M_{\rm SUSY}$
versus $m_{\rm LSP}$ for $\mu>0$, $A_0=0$,
$\Delta_{\tilde\tau_2}=1$, $\alpha_s(M_Z)=0.1185$ and
with $m_b(M_Z)=2.684~{\rm GeV}$ (dashed lines),
$3.092~{\rm GeV}$ (dotted lines) or $2.888~{\rm GeV}$
(solid lines). \label{figc}}
\end{figure}

\begin{itemize}
\item The lower bounds on $m_{\rm LSP}$ are not so
sensitive to the variations of $m_b(M_Z)$.
\item The lower bound on $m_{\rm LSP}$ from
Eq.~(\ref{mhb}) overshadows all others.
\item The upper bound on $m_{\rm LSP}$ from
Eq.~(\ref{cdmb}) is very sensitive to the
variations of $m_b(M_Z)$. In particular, one
notices the extreme sensitivity of the almost
vertical part of the corresponding line, where
the LSP annihilation via an $A$-boson exchange
in the $s$-channel is\cite{lah} by far the
dominant process, since $m_A$, which
is smaller than $2m_{\rm LSP}$, is always very
close to it as seen from Fig. \ref{figb}. This
sensitivity can be understood from Fig.~\ref{figc},
where $m_A$ is depicted versus $m_{\rm LSP}$ for
various $m_b(M_Z)$'s. We see that, as $m_b(M_Z)$
decreases, $m_A$ increases and approaches
$2m_{\rm LSP}$. The $A$-pole annihilation is then
enhanced and $\Omega_{\rm LSP}h^2$ is drastically
reduced causing an increase of the upper bound on
$m_{\rm LSP}$.
\item For $\Delta_{\tilde\tau_2}<0.25$, bino-stau
coannihilations\cite{ellis} take over leading to a
very pronounced reduction of $\Omega_{\rm LSP}h^2$,
thereby enhancing the upper limit on $m_{\rm LSP}$.

\end{itemize}

\par
For $\mu>0$, $\alpha_s(M_Z)=0.1185$ and $m_b(M_Z)=
2.888~{\rm GeV}$,
we find the following allowed ranges of parameters:
\bea && 176~{\rm GeV} \lesssim m_{\rm LSP}
\lesssim615~{\rm GeV},~~0\lesssim
\Delta_{\tilde\tau_2}\lesssim 1.8,
\nonumber \\
&& 58 \lesssim \tan\beta \lesssim
59,~~0.14\lesssim c\lesssim0.17.\eea

\section{The Inflationary Scenario}
\label{inflation}

One of the most promising inflationary scenarios is
hybrid inflation\cite{linde}, which uses two real
scalars: one which provides the vacuum energy for
inflation and a second which is the slowly varying
field during inflation. This scheme is naturally
incorporated\cite{hybrid} in SUSY GUTs, but in its
standard realization has the following
problem\cite{pana1}: if the GUT gauge symmetry
breaking predicts monopoles (and this is the case
of $G_{\rm PS}$), they are copiously produced at
the end of inflation leading to a cosmological
catastrophe\cite{kibble}. One way to remedy this
is to generate a shifted inflationary trajectory,
so that $G_{\rm PS}$ is already broken during
inflation. This could be achieved\cite{jean} in
our SUSY GUT model even before the introduction
of the extra Higgs superfields, but only by
utilizing non-renormalizable terms. However, the
introduction of $\phi$ and $\bar\phi$ very
naturally gives rise\cite{jean2} to a shifted
inflationary path with the use of renormalizable
interactions only.

\subsection{The Shifted Inflationary Path}
\label{inflationw}

\par
The superpotential terms which are relevant for
inflation are given by
\begin{equation}
W=\kappa S(H^c\bar{H}^c-M^2)-\beta S\phi^2+
m\bar{\phi}\phi+\lambda\bar{\phi}H^c\bar{H}^c,
\label{superpotential}
\end{equation}
where $M,m \sim M_{\rm GUT}\simeq2.86\times
10^{16}~{\rm GeV}$, and $\kappa$, $\beta$ and $\lambda$
are dimensionless coupling constants with
$M,~m,~\kappa,~\lambda>0$ by field redefinitions. For
simplicity, we take $\beta>0$. (The parameters are
normalized so that they correspond to the couplings
between the SM singlet components of the superfields.)

\par
The scalar potential obtained from $W$ is given by
\begin{eqnarray}
V=\left\vert\kappa(H^c\bar{H}^c-M^2)-\beta\phi^2
\right\vert^2+\left\vert 2\beta S\phi-m\bar{\phi}
\right\vert^2+\left\vert m\phi+\lambda H^c
\bar{H}^c\right\vert^2\nonumber \\ +\left\vert\kappa
S+\lambda\bar{\phi} \right\vert^2\left(\vert
H^c\vert^2+\vert\bar{H}^c \vert^2\right)+{\rm
D-terms}.~~~~~~~~~~~~~~~~ \label{potential}
\end{eqnarray}
Vanishing of the D-terms yields
$\bar{H}^c\,^{*}=e^{i\vartheta}H^c$ ($H^c$,
$\bar{H}^c$ lie in their right handed neutrino
directions). We restrict ourselves to the direction
with $\vartheta=0$
which contains the shifted inflationary path and the
SUSY vacua (see below). Performing appropriate R and
gauge transformations, we bring $S$, $H^c$ and
$\bar{H}^c$ to the positive real axis.

\par
From the potential in Eq.~(\ref{potential}), we find
that the SUSY vacuum lies at
\begin{equation}
\frac{H^c\bar{H}^c}{M^2}\equiv\left(\frac{v_0}
{M}\right)^2=\frac{1}{2\xi}\left(1-\sqrt{1-4\xi}\right),
~~\frac{\phi}{M}=-\sqrt{\frac{\kappa\xi}{\beta}}\left(
\frac{v_0}{M}\right)^2 \label{vacuum}
\end{equation}
with $S=0$ and $\bar{\phi}=0$, where
$\xi=\beta\lambda^2M^2/\kappa m^2<1/4$. The potential
possesses a shifted flat direction
(besides the trivial one) at
\begin{equation}
\frac{H^c\bar{H}^c}{M^2}\equiv\left(\frac{v}{M}
\right)^2=\frac{2\kappa^2(\frac{1}{4\xi}+1)+
\frac{\lambda^2}{\xi}}{2(\kappa^2+\lambda^2)},
~~\frac{\phi}{M}=-\frac{1}{2}\sqrt{\frac{\kappa}
{\beta\xi}},
~~\bar{\phi}=-\frac{\kappa}{\lambda}S
\label{trajectory}
\end{equation}
with $S>0$ and a constant potential energy density
$V_0$ given by
\begin{equation}
\frac{V_0}{M^4}=\frac{\kappa^2\lambda^2}{\kappa^2+
\lambda^2}\left(\frac{1}{4\xi}-1\right)^2,
\label{V0}
\end{equation}
which can be used as inflationary path. $V_0\neq0$
breaks SUSY on this path, while the constant non-zero
values of $H^c$, $\bar{H}^c$ break the GUT gauge
symmetry too. The SUSY breaking implies the existence
of one-loop radiative corrections\cite{DvaSha} which
lift the classical flatness of this path, yielding
the necessary inclination for driving the
inflaton towards the SUSY vacuum.

\par
The one-loop radiative corrections to $V$ along the
shifted inflationary trajectory are calculated by
using the Coleman-Weinberg formula\cite{cw}:
\begin{equation}
\Delta V=\frac{1}{64\pi^2}\sum_i(-)^{F_i}M_i^4\ln
\frac{M_i^2}{\Lambda^2}, \label{Coleman}
\end{equation}
where the sum extends over all helicity states $i$,
$F_i$ and $M_i^2$ are the fermion number and mass
squared of the $i$th state and $\Lambda$ is a
renormalization mass scale. In order
to use this formula for creating a logarithmic
slope which drives the canonically normalized real
inflaton field
$\sigma=\sqrt{2(\kappa^2+\lambda^2)}S/\lambda$
towards the minimum, one has first to derive the
mass spectrum of the model on the shifted
inflationary path. This is a quite complicated
task and we will skip it here.

\subsection{Inflationary Observables}
\label{inflationq}

The slow roll parameters are given by (see e.g.
Ref.~\refcite{lectures})
\begin{equation}
\varepsilon\simeq\frac{m_{\rm P}^2}{2}~\left(
\frac{V'(\sigma)}{V_0}\right)^2, ~~~~~~~~\eta
\simeq m_{\rm
P}^2~\frac{V''(\sigma)} {V_0}, \label{slowroll}
\end{equation}
where the primes denote derivation with respect to
the real normalized inflaton field $\sigma$ and
$m_{\rm P}\simeq 2.44\times 10^{18}~{\rm GeV}$ is
the reduced Planck scale. The conditions for
inflation to take place are $\varepsilon\leq 1$ and
$\vert\eta\vert\leq 1$.

\par
Calculating the number of e-foldings $N_Q$ that our
present horizon scale suffered during inflation, we
obtain the following relation (see e.g.
Ref.~\refcite{lectures}):
\begin{equation}
N_Q\simeq\frac{1}{m_{\rm P}^2}\int_{\sigma_f}
^{\sigma_Q}\frac{V_0}{V'(\sigma)}d\sigma\simeq
\ln\left(4.41\times 10^{11}~T_r^{\frac{1}{3}}~
V_0^{\frac{1}{6}}\right), \label{NQ}
\end{equation}
where $\sigma_f~[\sigma_Q]$ is the value of $\sigma$ at
the end of inflation [when our present horizon scale
crossed outside the inflationary horizon] and
$T_r\simeq 10^9~{\rm GeV}$ is the reheat temperature
taken to saturate the gravitino
constraint\cite{gravitino}.

\par
The quadrupole anisotropy of the cosmic microwave
background radiation can be calculated as follows (see
e.g. Ref.~\refcite{lectures}):
\begin{equation}
\left(\frac{\delta T}{T}\right)_Q\simeq\frac{1}
{12\sqrt{5}}\frac{V_0^{\frac{3}{2}}} {V'(\sigma_Q)m_{\rm
P}^3}\cdot \label{anisotropy}
\end{equation}
Fixing $(\delta T/T)_Q\simeq 6.6\times 10^{-6}$, which is
its central value from the cosmic background explorer
(COBE)\cite{cobe} (assuming that the spectral index $n=1$),
we can determine one of the free parameters (say $\beta$)
in terms of the others ($m$, $\kappa$ and $\lambda$). For
instance, we find $\beta=0.1$, for $m=4.35\times 10^{15}~
{\rm GeV}$ and $\kappa=\lambda= 3\times 10^{-2}$. In this
case, the instability point of the shifted path lies at
$\sigma_c\simeq 3.55\times 10^{16}~{\rm GeV}$, $\sigma_f
\simeq 1.7\times 10^{17}~{\rm GeV}$ and $\sigma_Q \simeq
1.6\times 10^{18}~{\rm GeV}$\cite{lectures}. Also,
$M\simeq 2.66\times 10^{16}~{\rm GeV}$, $N_Q\simeq 57.7$
and $n\simeq 0.98$. Note that the slow roll conditions
are violated and, thus, inflation ends well before
reaching the instability point at $\sigma_c$. We see that
the COBE constraint can be easily satisfied with natural
values of the parameters. Moreover, superheavy SM
non-singlets with masses $\ll M_{\rm GUT}$, which could
disturb the unification of the MSSM gauge coupling
constants, are not encountered.

\subsection{Supergravity Corrections} \label{sugra}

Here, inflation takes place at quite high $\sigma$'s.
So, supergravity (SUGRA) corrections are important
and could easily invalidate inflation, in contrast
to the standard hybrid inflation case, where they
can be kept\cite{inf} under control. This catastrophe
can be avoided by invoking a specific K\"ahler
potential (used in no-scale SUGRA models) and a gauge
singlet $Z$ with a similar K\"ahler potential, as is
suggested in Ref.~\refcite{pana2}. Assuming a
superheavy VEV for $Z$ via D-terms, one can achieve
an exact cancellation of the inflaton mass corrections
on the shifted path. So, inflation remains intact, but
gets considerably corrected via the kinetic terms of
$\sigma$.

\par
We find that, for the $\sigma$'s under consideration,
the SUGRA corrections have only a small influence on
$\sigma_Q$ if we use the same input values for the
free parameters as in the global SUSY case. On the
contrary, $(\delta T/T)_Q$ increases considerably.
However, we can easily readjust the parameters so
that the COBE constraint is again met. For instance,
$(\delta T/T)_Q\simeq 6.6\times 10^{-6}$ is now
obtained with $m=3.8\times 10^{15}~{\rm GeV}$ keeping
$\kappa=\lambda =3\times 10^{-2}$, $\beta=0.1$ as in
global SUSY. In this case, $\sigma_c\simeq 2.7\times
10^{16}~{\rm GeV}$, $\sigma_f\simeq 1.8\times
10^{17}~{\rm GeV}$ and $\sigma_Q\simeq 1.6\times
10^{18}~{\rm GeV}$. Also, $M\simeq 2.6\times
10^{16}~{\rm GeV}$, $N_Q\simeq 57.5$ and $n\simeq
0.99$.

\section{Conclusions} \label{conclusions}

We have reviewed the construction of a SUSY GUT model
based on the PS gauge group which naturally yields a
YQUC, allowing an acceptable $b$-quark mass within the
CMSSM with $\mu>0$. We found that there exists a wide
and natural range of parameters consistent
with the data on the CDM abundance in the universe,
$b\rightarrow s\gamma$, the muon anomalous magnetic
moment and the Higgs boson masses. Moreover, the model
gives rise to a new version of the
shifted hybrid inflationary scenario, which avoids
overproduction of monopoles at the end of
inflation by using only renormalizable interactions.

\section*{Acknowledgments}

We would like to thank M.E. G\'omez, R. Jeannerot and
S. Khalil for fruitful and pleasant collaborations from
which this work is culled. This work was supported by
European Union under the RTN contracts HPRN-CT-2000-00148
and HPRN-CT-2000-00152.

\def\ijmp#1#2#3{{Int. Jour. Mod. Phys.}
{\bf #1},~#3~(#2)}
\def\plb#1#2#3{{Phys. Lett. B }{\bf #1},~#3~(#2)}
\def\zpc#1#2#3{{Z. Phys. C }{\bf #1},~#3~(#2)}
\def\prl#1#2#3{{Phys. Rev. Lett.}
{\bf #1},~#3~(#2)}
\def\rmp#1#2#3{{Rev. Mod. Phys.}
{\bf #1},~#3~(#2)}
\def\prep#1#2#3{{Phys. Rep. }{\bf #1},~#3~(#2)}
\def\prd#1#2#3{{Phys. Rev. D }{\bf #1},~#3~(#2)}
\def\npb#1#2#3{{Nucl. Phys. }{\bf B#1},~#3~(#2)}
\def\npps#1#2#3{{Nucl. Phys. B (Proc. Sup.)}
{\bf #1},~#3~(#2)}
\def\mpl#1#2#3{{Mod. Phys. Lett.}
{\bf #1},~#3~(#2)}
\def\arnps#1#2#3{{Annu. Rev. Nucl. Part. Sci.}
{\bf #1},~#3~(#2)}
\def\sjnp#1#2#3{{Sov. J. Nucl. Phys.}
{\bf #1},~#3~(#2)}
\def\jetp#1#2#3{{JETP Lett. }{\bf #1},~#3~(#2)}
\def\app#1#2#3{{Acta Phys. Polon.}
{\bf #1},~#3~(#2)}
\def\rnc#1#2#3{{Riv. Nuovo Cim.}
{\bf #1},~#3~(#2)}
\def\ap#1#2#3{{Ann. Phys. }{\bf #1},~#3~(#2)}
\def\ptp#1#2#3{{Prog. Theor. Phys.}
{\bf #1},~#3~(#2)}
\def\apjl#1#2#3{{Astrophys. J. Lett.}
{\bf #1},~#3~(#2)}
\def\n#1#2#3{{Nature }{\bf #1},~#3~(#2)}
\def\apj#1#2#3{{Astrophys. J.}
{\bf #1},~#3~(#2)}
\def\anj#1#2#3{{Astron. J. }{\bf #1},~#3~(#2)}
\def\mnras#1#2#3{{MNRAS }{\bf #1},~#3~(#2)}
\def\grg#1#2#3{{Gen. Rel. Grav.}
{\bf #1},~#3~(#2)}
\def\s#1#2#3{{Science }{\bf #1},~#3~(#2)}
\def\baas#1#2#3{{Bull. Am. Astron. Soc.}
{\bf #1},~#3~(#2)}
\def\ibid#1#2#3{{\it ibid. }{\bf #1},~#3~(#2)}
\def\cpc#1#2#3{{Comput. Phys. Commun.}
{\bf #1},~#3~(#2)}
\def\astp#1#2#3{{Astropart. Phys.}
{\bf #1},~#3~(#2)}
\def\epjc#1#2#3{{Eur. Phys. J. C}
{\bf #1},~#3~(#2)}
\def\nima#1#2#3{{Nucl. Instrum. Meth. A}
{\bf #1},~#3~(#2)}
\def\jhep#1#2#3{{J. High Energy Phys.}
{\bf #1},~#3~(#2)}

\end{document}